# Direct measurement and modelling of internal strains in ion-implanted diamond

F. Bosia[1,2,3(*)], N. Argiolas[3,4], M. Bazzan[3,4], B. A. Fairchild[5], A. D. Greentree[6], D. W. M. Lau[5], P. Olivero[1,2,3], F. Picollo[1,2,3], S. Rubanov[5], S. Prawer[5]

[1]*Department of Physics - NIS Centre of Excellence, Università di Torino, Italy*

[2]*Istituto Nazionale di Fisica Nucleare (INFN), Sezione di Torino, Italy*

[3]*Consorzio Nazionale Interuniversitario per le Scienze Fisiche della Materia (CNISM)*

[4]*Physics and Astronomy Department, Università di Padova, Italy*

[5]*School of Physics, The University of Melbourne, Melbourne 3010, Australia*

[6]*Applied Physics, School of Applied Sciences, RMIT University, Melbourne 3001, Australia*

(*) Corresponding author: federico.bosia@unito.it

## Abstract

We present a phenomenological model and Finite Element simulations to describe the depth variation of mass density and strain of ion-implanted single-crystal diamond. Several experiments are employed to validate the approach: firstly, samples implanted with 180 keV B ions at relatively low fluences are characterized using high-resolution X-ray diffraction (HR-XRD); secondly, the mass density variation of a sample implanted with 500 keV He ions well above its amorphization threshold is characterized with Electron Energy Loss Spectroscopy (EELS). At high damage densities, the experimental depth profiles of strain and density display a saturation effect with increasing damage and a shift of the damage density peak towards greater depth values with respect to those predicted by TRIM simulations, which are well accounted for in the model presented here. The model is then further validated by comparing TEM-measured and simulated thickness values of a buried amorphous carbon layer formed at different depths by implantation of 500 keV He ions through a variable-thickness mask to simulate the simultaneous implantation of ions at different energies.

## 1.        Introduction

Ion implantation has been widely applied to the fabrication and functionalization of single crystal diamond, with application in diverse fields such as optics and photonics[1-8], bio-sensors[9], particle detectors[10, 11] and micro-electromechanical systems (MEMS) [12-14]. Several fabrication schemes can be implemented by exploiting light MeV ions, whose strongly non-uniform damage depth profile allows the creation of heavily damaged buried layers which graphitize after thermal annealing, whilst the structure of the surrounding material is largely restored[15-17]. Thus, spatially well-defined structures can be created by selectively etching the graphitized regions[18] or graphitic conductive paths can be fabricated for specific applications[19, 20].

        To reliably design and fabricate structures with MeV ion beam lithography, accurate control of the spatial extension of the graphitized layer is necessary.  However, the mechanisms by which the diamond lattice structure is modified by ion-beam processes are still not fully understood.  This is due to the complex interplay of the various parameters involved, including implanted ion species and energy, implantation fluence and temperature, post-implantation annealing temperature, local stress, and even fundamental material properties of diamond such as tensile strength are still poorly understood[21]. It has been shown[22] that primary defects, formed in the collision cascades during ion implantation, consist primarily of vacancy and interstitial atoms. At room temperature, vacancies are immobile, but interstitials can diffuse substantially, either recombining with vacancies or moving out of the implanted region[22].

        One concept of particular importance to ion-beam modification of diamond is that of the critical density for amorphisation, $D_C$. This parameter is typically referred to as the

damage level (quantified as a vacancy concentration via TRIM modelling[23]) beyond which the diamond lattice is permanently amorphised, and subsequently graphitizes upon thermal annealing[24]. Despite the central role that $D_C$ plays in the ion-beam modification of diamond, considerable uncertainty remains on the value of $D_C$ and its dependence on implantation parameters (e.g. depth and/or local strain, self-annealing, etc)[15, 16, 25-33]. Recently it was shown that $D_C$ appears to be set by local density and tensile strain, rather than simply by vacancy concentration, and a threshold density for amorphisation of approximately 2.95 g/cm$^3$, corresponding to a strain value of approximately 16% was found[34].

Another extremely important issue when designing implantation strategies is that of damage accumulation, and in particular vacancy saturation[35]. As ions are implanted into the sample, they cause damage, and hence the properties of the sample vary as a function of the accumulated dose. The result is that the damage per implanted ion reduces as a function of the local damage, leading to a marked departure of the inferred vacancy concentration from the predictions of standard Monte-Carlo simulation packages such as TRIM[23], which typically models ion-implantation into *virgin* material. Interestingly, saturation behaviour is also observed in both the $sp^2$ fraction and density with increasing damage, although the point of saturation is *beyond* the amorphisation threshold, indicating continued modification of the sample with increasing damage even after amorphisation has been achieved. [34]

Another interesting feature of amorphous carbons that is related to vacancy saturation is that high fluence ion-implantation tends to create amorphous carbon with a density around 2.2 g/cm$^3$. In particular, it is observed that high-density carbons (diamond[34], $\rho_d$ = 3.52 g/cm$^3$ and tetrahedral amorphous carbon[36], $\rho_{taC}$ = 3.3 g/cm$^3$) expand with increasing fluence, whereas low density glassy carbon[37], $\rho_{gC}$ = 1.55 g/cm$^3$ compactifies. This limit for ion-beam damaged carbon materials will be used in our modelling below.

The highly non-trivial nature of the ion-induced damage of diamond means that to accurately predict (or equivalently design) an implantation strategy, sophisticated modelling is required. Such modelling must include geometrical parameters, to correctly include the mechanical constraining effects of the surrounding undamaged diamond material and variation in elastic constants of the implanted substrate. Our work discusses the generation of such models.

Here we present a numerical procedure, based on a phenomenological model for damage accumulation and saturation and subsequent use of Finite Element (FEM) simulations, which allows the reliable determination of strains in an ion-implanted diamond substrate. This model is based on previous studies by some of the authors[38, 39] and represents their completion. We validate the FEM modelling approach in the case of samples implanted at low fluences, where damage saturation effects are negligible, by comparing numerical results with experimental High-Resolution X-Ray Diffraction (HR-XRD) measurements. Subsequently, we check the predictions of the whole numerical procedure, including saturation effects, using Electronic Energy Loss Spectroscopy (EELS) and Transmission Electron Microscopy (TEM) experimental data from samples implanted at high damage densities.

The paper is organized as follows: in Section 2 the model is outlined; in Sections 3 and 4 the model predictions are compared to experimental results in samples implanted respectively in low and high damage regimes.

## 2. The Model

In the present study, we adopt and extend the model presented in [38], which accounts for saturation in the creation of vacancies in the damaged diamond crystal lattice at increasing implantation fluences. The model is inspired by the work reported in [35] and takes into account the concentration of ion-induced vacancies with a simple linear approximation for the probability for a newly created vacancy to recombine with a self-interstitial:

$$P_{rec}(F,z) = \frac{\rho_V(F,z)}{\alpha},$$
(1)

where $P_{rec}$ is the recombination probability at a given depth $z$ and implantation fluence $F$, $\rho_V$ is the vacancy density in the material at depth $z$ and $\alpha$ is an empirical parameter depending on the implantation conditions that accounts for the defect recombination probability. In particular, $\alpha$ represents the saturation vacancy density where the probability of recombination is 1 and therefore no further vacancies are introduced into the structure by additional implanted ions. In other words when $\rho_V = \alpha$, the energy deposited by each additional ion at that depth simply moves atoms, without creating additional vacancies. By solving the associated differential equation

$$\frac{d\rho_V}{dF} = \left(1 - \frac{\rho_V}{\alpha}\right)\lambda(z),    (2)$$

where $\lambda(z)$ is the linear vacancy depth profile calculated using the TRIM 2008.04 code[23], we obtain the actual vacancy density of the damaged diamond as a function of substrate depth and implantation fluence:

$$\rho_V(F,z) = \alpha\left(1 - e^{-\frac{\lambda(z)F}{\alpha}}\right)    (3)$$

By assuming that the mass density $\rho$ of the damaged diamond is linearly proportional to the vacancy density and considering the known boundary values of $\rho$ at zero and infinite fluences, we obtain the following expression:

$$\rho(F,z) = \rho_d - \left(\rho_d - \rho_{aC}\right)\left(1 - e^{-\frac{\lambda(z)F}{\alpha}}\right)    (4)$$

where $\rho_d = 3.52$ g cm$^{-3}$ is the diamond density and $\rho_{aC} = 2.06$ g cm$^{-3}$ is the limiting density of the ion-damaged material, as determined in [34].

In addition to vacancy saturation, the reduction in density of the material leads perforce to swelling. As the diamond below the implantation is undamaged, this swelling must manifest on the front surface (implantation side) of the diamond. We observe this swelling as a shift of the damage peak towards *greater* depth values, measured with respect to the implanted surface, than those predicted by the TRIM code, as the latter calculates ion trajectories into a pristine substrate. Furthermore, the ions experience a reduced material density with increasing accumulated damage, so that the mean implantation depth increases as the material progressively swells. This can be accounted for by introducing a rescaling of the depth coordinate $z$ due to the reduced stopping power of the substrate, which is proportional to the mass density decrease in the damaged layer. To numerically account for this effect, we thus divide the depth coordinate in a large number ($N$) of intervals of width $\Delta z$, and rescale the $i$-th depth as:

$$z_i \rightarrow z'_i = z'_{i-1} + \frac{\rho_d}{\rho(z_i)} \Delta z \qquad (5)$$

where $i=1...N$ is the interval label, $z$ is the original coordinate, $z'$ is the rescaled coordinate, and $\rho(z_i)$ the implanted diamond density at depth $z_i$. Thus, the piecewise-defined rescaled mass density distribution $\rho(z')$ can be determined. As reported in [17], following a "rule of mixture" approach, we assume for simplicity that the mechanical parameters of the damaged material (such as the Young's modulus $E$ and the Poisson's ratio $\nu$) have the same functional variation as that of mass density:

$$E(F, z') = E_d - \left(E_d - E_{aC}\right)\left(1 - e^{-\frac{\lambda(z')F}{\alpha}}\right) \qquad (6a)$$

$$\nu(F, z') = \nu_d - \left(\nu_d - \nu_{aC}\right)\left(1 - e^{-\frac{\lambda(z')F}{\alpha}}\right) \qquad (6b)$$

with $E_d = 1220$ GPa, $E_{aC} = 10$ GPa, $\nu_d = 0.2$ and $\nu_{aC} = 0.18$[40].

Once the ion species and energy as well as the implantation fluence and geometry are known, one can perform a standard TRIM simulation to obtain the vacancy density profile. All TRIM simulations were performed in "Detailed calculation with full damage cascade" mode, by imposing a displacement energy for carbon atoms in diamond of 50 eV[41]. The TRIM output is used in Eqs. 4 - 6 to calculate the spatial variation of both structural and mechanical parameters of the irradiated sample. These data are then fed into a 2-D or 3-D FEM model to simulate strains, stresses and surface deformations occurring in the diamond substrate as a result of ion implantation. In our work, simulations are carried out using the "Structural mechanics" module of COMSOL Multiphysics®[42]. As explained in[38], the simulations are performed by imposing a constrained isotropic volume expansion in the damaged regions that is proportional to the local density variation and affected by the local variation of the mechanical properties.

### 3.   Low damage density regime

To first evaluate the predictive abilities of FEM modelling only, we considered samples implanted at relatively low fluences, to exclude the damage saturation effects described above.

#### 3.1     Samples and ion implantation

Ion implantation was performed on optical-grade CVD single-crystal samples produced by Element Six. The samples are classified as type IIa (single substitutional nitrogen concentration below 1 ppm, single substitutional boron concentration below 50 ppb), have 100 crystal orientation and are $3 \times 3 \times 0.5$ mm$^3$ in size, with two optically polished opposite large faces. To be suitably analyzed with the HD-XRD technique, diamond samples with relatively shallow ion implantations are required, thus allowing the observation of sufficiently resolved thickness fringes and the reconstruction of the damage profile from simulations. We implanted the samples at room temperature with 180 keV B ions at the Olivetti I-Jet facilities (Arnad, Italy). The projected range for this implantation is 268 nm with longitudinal straggle 45 nm. Three samples were prepared with uniform irradiation of the whole upper surface at fluences of $5 \times 10^{13}$ cm$^{-2}$, $1 \times 10^{14}$ cm$^{-2}$ and $5 \times 10^{14}$ cm$^{-2}$.

### 3.2    Experimental measurements

The B-implanted samples were investigated with HR-XRD (details of the experimental technique and data analysis can be found in[43]) at the department of Physics and Astronomy of the University of Padova by means of a Philips MRD diffractometer. The source was an X-ray tube with copper anode, equipped with a parabolic mirror and a Ge (2 2 0) four-bounce Bartels monochromator. The resulting primary beam had a divergence angle of 0.0039° in the equatorial plane and a spectral purity of $\Delta\lambda / \lambda = 10^{-5}$ at a wavelength $\lambda = 1.54056$ Å. The beam impinges on the sample at an angle $\omega$, measured by a high precision goniometer, on which the specimen is mounted. The scattered radiation from the sample was measured as a function of the incidence angle $\omega$ and the scattering angle $2\theta$ by a Xe proportional detector mounted on a second independent goniometer, coaxial to the first. To improve the angular resolution of the measurement, the detector was equipped with a three-bounce Ge (2 2 0) analyzer that guarantees an acceptance angle of 0.0039°. The system was maintained in a measuring chamber at a constant temperature of $(25.0 \pm 0.1)$ °C.

Both unimplanted and implanted samples were measured under the same conditions near the (0 0 4) reflection using a combination of radial $\omega$-$2\theta$ scans and reciprocal space maps. Moreover, the presence of interface relaxation was checked by acquiring asymmetric reciprocal space maps around the (3 1 1) reflection. The unimplanted sample revealed a strong Bragg peak corresponding to the (0 0 4) lattice planes at the Bragg angle $\theta_B = (59.746 \pm 0.001)°$ from which a lattice parameter $a = 3.567$ Å can be determined, in good agreement with the literature[44]. However, the measured width of the Bragg peak was $(6.43 \pm 0.004)° \cdot 10^{-3}$ which is significantly larger than the theoretical width $2.68° \cdot 10^{-3}$ calculated using dynamical diffraction theory and corrected for experimental broadening[43]. This observation indicates that some kind of distortion is natively present in the substrate. By reducing the beam footprint to few tens of μm with the aid of some collimation slits and performing a series of $\omega$ – scans in different positions of the sample, we observed a slowly-varying bending of the substrate, with a minimal curvature radius of 32 m, probably due to residual stresses from the growth process. As this variation occurs over distances much larger than the coherence length of the X-ray wavefield (i.e. hundreds of nanometers) this effect can simply be modelled considering the measured signal as the superposition of many diffraction spectra coming from slightly tilted samples.

The analysis of the asymmetrical maps (not shown here) did not reveal the presence of interface relaxation in any of the implanted samples. In the following, it will therefore be assumed that the deformed layers are fully pseudomorphic with the substrate, so that all near-surface deformations occur along the vertical direction. An example of one of the reciprocal space maps obtained from the sample implanted at fluence $5 \times 10^{14}$ cm$^{-2}$ around the symmetric (0 0 4) reflection is shown in Fig. 1. The maps were projected along the $Q_z$ direction to obtain the usual $\omega$-$2\theta$ scan, but with a higher accuracy with respect to simple line-scan measurements (Fig. 2).

The curves were analysed using an in-house developed program that simulates diffraction curves produced by a deformed sample using the Takagi - Taupin dynamical diffraction theory[43]. The *deformation profile* $\Delta d/d(z)$ (i.e. the relative change in the vertical crystal lattice parameter, due to the local defect concentration, as a function of depth) was parametrized using a $10^{\text{th}}$ order spline [45], so that each profile was described by 10 parameters, as well as a further one to define the depth of the profile by rescaling abscissae of the spline curve. This profile was then discretized by a numerical routine into a series of $N$ layers, each characterized by a local $\Delta d_i/d$ ($i = 1...N$), to allow the calculation of the diffraction spectrum in the framework of the Takagi – Taupin approach[43]. To account for the gradual disruption of crystalline order introduced in the crystal structure by ion implantation, a "*disorder profile*" was also included in the model [43]. Structural disorder was modelled as a random displacement of the atoms around their average lattice position. For the sake of simplicity, the displacements were assumed to be isotropically distributed according to a zero-mean Gaussian distribution, producing an effect similar to that of thermal agitation on the diffracted field. The parameter used to characterize this effect was the local root mean square of displacements from equilibrium positions. To reduce the number of fitting parameters, the disorder profile was assumed to be proportional to the deformation profile, thus obtaining a reasonable compromise between accuracy of the fit and number of fitting parameters. An additional parameter relating the disorder profile to the deformation one is therefore included in the fitting procedure. The effect of the above-mentioned macroscopic substrate distortion due to residual growth stresses was described by convolving the simulated profile with a broadening Gaussian function, whose width is chosen so that the width of the simulated substrate peak width coincides with the experimental one. The parameters defining the deformation profile are finally obtained by minimizing the function $\chi^2_{log} = \sum_i \big[ \log(I_i) - \log\big(f(\theta_i)\big) \big]^2$, where $\theta_i$ is the scattering angle, $f(\theta_i)$ is the diffraction curve intensity in $\theta_i$ calculated using the simulation code and $I_i$ is the intensity measured in $\theta_i$. The use of the logarithmic function allows balanced weighing of points on the curve with marked differences in intensity. An example of experimental and simulated rocking curves for the sample implanted at fluence $5 \times 10^{14}$ cm$^{-2}$ is shown in Fig. 2 with good agreement between experimental and simulated data.

The resulting deformation profile $\Delta d/d(z)$ for the three samples is shown in Fig. 3a. The profiles are normalized with respect to an arbitrary factor proportional to their implantation fluence for comparison purposes in Fig. 3b (specifically, the profiles relative to fluences of $5 \cdot 10^{13}$ cm$^{-2}$, $10^{14}$ and $5 \cdot 10^{14}$ cm$^{-2}$ are divided by the factors 0.0025, 0.005 and 0.025, respectively). The three curves are in good agreement, show decreasing $\Delta d/d(z)$ with increasing fluence indicating the onset of saturation effects. Similar curves to those in Fig. 3a are obtained for the calculated disorder (not shown here).

According to the reciprocal space mapping analysis performed near the (3 1 1) asymmetric reflection, there is no lateral relaxation of the implanted layer, so the deformation profile measured by HR-XRD derives from two effects: the ion-induced change in the crystal density and the vertical strain caused by the Poisson effect. The $\varepsilon_z$ principal strain component can therefore be extracted from the measured data through the following relation:

$$\varepsilon_z\left(F,z\right)=\frac{\Delta d}{d}\left(F,z\right)-\frac{\Delta\rho}{\rho}\left(F,z\right)=\frac{\Delta d}{d}\left(F,z\right)+\frac{\rho_d-\rho_{aC}}{\rho_d}\cdot\left(1-e^{\frac{F\lambda(z)}{\alpha}}\right) \quad (7)$$

The data calculated with this relation can therefore be compared to simulated strain values, as discussed in the following section.

### 3.3    Data analysis and strain modelling

Due to the low value of irradiation fluence, the effects of damage saturation and swelling are minimal for the samples considered in this Section. Simulations to determine vacancy densities were carried out using the Crystal-TRIM (C-TRIM) code[46], which accounts for local electronic energy loss using a semi-empirical formula based on a modified Oen-Robinson model[46], and has particularly well documented simulation parameters for B implantations. The C-TRIM-derived linear vacancy depth profile $\lambda(z)$ is shown in Fig. 3b together with the experimental HR-XRD data. This curve is in good agreement with the experimental ones, confirming that the deformation occurring in the samples is indeed predominantly vacancy related, and that corrections for damage saturation and swelling are relevant only for higher damage density regimes.

As mentioned in the previous Section, FEM simulations were performed to determine the strain profile in the implanted samples, and were carried out using as input the vacancy density profile resulting from C-TRIM. The resulting tensile strain profile for the sample implanted at a fluence of $5 \times 10^{14} cm^{-2}$ is shown in Fig. 4, together with the experimental values determined from the HR-XRD data shown in Fig. 3 using Eq. 6. An excellent agreement between the HR-XRD data and the numerical results is observed in Fig. 4 and also for the other two samples (not shown here), both in amplitude and depth distribution, proving the validity of the adopted procedure. As expected, in this case the strain levels are quite low (at most 1.5%), due to the uniform irradiation of the whole of the specimen surface at relatively low fluences, and well below the 16% value indicated in [34] as the threshold for amorphization.

## 4       High damage density regime

Higher damage densities were investigated, corresponding to higher strain levels, to validate the damage-saturation analytical model presented in Section 2.

### *4.1     Samples and ion implantation*

For the higher density experiments, we used a single-crystal HPHT diamond produced by Sumitomo Electrics. The sample is classified as type Ib (single substitutional nitrogen concentration 10-100 ppm), has 100 crystal orientation and is $3 \times 3 \times 1.5 mm^3$ in size, with two optically polished opposite large faces. The sample was implanted at room temperature with 500 keV He ions with a raster-scanning ion microbeam of the 5U NEC Pelletron accelerator of the University of Melbourne. The implantation fluence was $5 \times 10^{16} cm^{-2}$, the sample was tilted by 3° to reduce channelling effects. A dual beam focused ion beam (FIB) system was used to prepare thin cross-sections in the [011] orientation for TEM and EELS measurements, using a standard FIB liftout process[47].

### 4.2 EELS measurements

The TEM bright field imaging was performed using a Tecnai TF20 electron microscope operating at 200 kV terminal voltage. EELS analysis was performed on a JEOL 2100F TEM equipped with a Gatan Tridium spectrometer. The low loss spectrum was collected in scanning TEM (STEM) mode with a convergence semi-angle of 9.5 mrad and a collection angle of 3.5 mrad. A series of line scans with a probe size of 0.2 nm was rastered across the sample in 0.2-1 nm steps, with a collection time of 1 s per spectrum.

Density was determined using a method based on the plasmon energy in the low-loss region of the EELS spectrum[48], assuming 4 valence electrons and an effective mass of the electron $m_{e,eff} = 0.84 \times m_e$. The $\pi^*$ contribution was calculated by integrating the intensity under the feature at ~284.5 eV while the $\sigma^*$ contribution was calculated by integrating the intensity 30 eV from the edge onset.

Low-loss EELS spectra were collected across several line scans with increasing depth of the sample. The EELS data were processed to fit diamond, amorphous carbon, $\pi^*$ and $\sigma^*$ peaks, to determine the density and ratio of $\pi^*$ to $\sigma^*$ within the sample. Results are discussed in the next Section, together with numerical data.

### 4.3 EELS data analysis and strain modelling

TRIM simulations for 500 keV He ions in diamond were performed since no reliable parameters are available in the literature for the implementation of C-TRIM simulations with the above-mentioned ion species. Subsequently, the analytic procedure described in Section 2 was applied to the data to account for damage saturation and swelling effects. Figure 5a illustrates how the TRIM vacancy density profile (continuous line) is modified in amplitude (dashed line) and in depth (dotted line) when damage saturation and swelling processes are taken into account by our model. A value of $\alpha = 7 \times 10^{22}$ cm$^{-3}$ was used, as determined from previous studies on similar implantations[39]. The resulting mass density values are shown in Fig. 5b, together with experimental EELS data (continuous line). The mass density profile is non-uniform and falls gradually on the leading edge of the amorphous carbon (a-C) region, to a minimum value of 2.06 g cm$^{-3}$ [36,37] before rising sharply at the trailing edge of the a-C zone. As discussed in [34], below a density of approximately 2.95 g cm$^{-3}$, diamond collapses into amorphous carbon because the strain is so great it is released by a structural transformation to a lower energy state. Using the given parameters, the agreement between experimental and numerical values is very satisfactory as regards to the peak amplitude and depth. Experimental data however display a more abrupt variation, compared to calculated values, from pristine diamond density (3.52 g cm$^{-3}$) to an heavily distorted diamond region at about $z = 0.9$ μm and back to diamond at about $z = 1.2$ μm (with an intermediate a-C region between $z = 0.93$ and $z = 1.12$ μm). We attribute this discrepancy to in situ defect self-annealing mechanisms, since implantations were carried out at room temperature and the current density used in the microbeam was about an order of magnitude greater than that used in implantations in Section 3. Thus, during the 500 keV He implantation in lightly damaged regions close to the surface, the self-annealing effect enhances defect recombination and induces the crystal to further revert to its pristine state, with negligible mass density variation. This is no longer true for vacancy density values above a threshold density $\rho_{th}$, estimated here as $4 \times 10^{22}$ cm$^{-3}$ (i.e. smaller than $D_C$), which occurs at $z_{th1} = 0.9$ μm and $z_{th2} = 1.2$ μm. As seen elsewhere[49], both the highly distorted diamond and the a-C that are present in this region convert to graphite upon high-temperature annealing. To account for this, an additional "graphitization threshold" was introduced in Eq. (4) and the theoretical mass density profile modified accordingly, as follows:

$$\rho(z) = \rho_d - (\rho_d - \rho_{aC}) \left(1 - e^{-\frac{\lambda(z)F}{\alpha}}\right) \frac{e^{\frac{(z-z_{th2})}{\beta}}}{\left(1 + e^{\frac{(z-z_{th1})}{\beta}}\right)\left(1 + e^{\frac{(z-z_{th2})}{\beta}}\right)} \tag{8}$$

where $\beta$ is a fitting constant that depends on the annealing temperature and determines the "steepness" of the threshold transition (a step-like transition is obtained at an annealing temperature of approximately 1400 deg. C). For room temperature implantations, $\beta = 10^{-8}$ m. The corresponding (long-dash) curve is shown in Fig. 5b, now providing an excellent agreement with experimental data. This mass density variation was used in FEM simulations to determine strains and surface swelling in the considered specimen.

In [34], a strain value of 16% is reported as the amorphization threshold of diamond, as derived from the relative density variation. A more precise estimation of this value and of the corresponding stresses can be sought through FEM simulations. The strains were calculated at the centre of the implanted area ($x = y = 0$ μm) and the following components are obtained at the amorphization threshold $Dc$, occurring at a depth of $z = 0.93$ μm[34]: $\varepsilon_x = \varepsilon_y = 0.05\%$, $\varepsilon_z = 18\%$. Under the assumptions of Eq. (6), we can determine the corresponding stress state as $\sigma_x = \sigma_y = -40.5$ GPa, $\sigma_z = 42.7$ GPa, which are much lower values than the upper bounds reported in [34]. The spatial variation of the calculated principal (i.e. along the *z* axis) component of the strain field in the depth direction, $\varepsilon_z(x, z)$, is shown in Fig. 6a. It is apparent that the strain profile closely follows that of the vacancy / mass density variation, with additional strain concentrations at the edges of the implanted area. The internal strains are responsible for the swelling effect at the surface, also shown in the figure (vertical displacements are magnified by a factor 50, to highlight the swelling effect).

We performed profilometry on the implanted sample to determine the swelling due to the implantation, and compared it with numerically predicted values. Profiles were obtained using an Ambios XP stylus profiler operating at a speed of 0.01 mm s$^{-1}$ and force = 0.05 mg. For each run the system was calibrated using a 186.2 nm vertical standard. A typical surface roughness of the sample before implantation was ~5 nm. A typical profilometric result is shown in Fig. 6b, together with the corresponding numerical profile derived from our model. The experimentally measured surface morphology displays some irregularities and noise of the order of ~5-10 nm, and the numerically-calculated swelling decreases more sharply at the borders of the implanted area. This smoother roll off of the experimental results is due to the ion straggling effects and the Gaussian shape of the ion microbeam, which are not accounted for in the FEM simulations. Apart from these edge effects, the experimental and numerical datasets are in good agreement.

### 4.4 *TEM measurements on a sample with an emerging damaged layer*

As further confirmation of the predictive capabilities of the adopted model, a further irradiation configuration was analysed. The sample under analysis was a type Ib HPHT single-crystal diamond (Sumitomo Electrics). As for the previous case, the sample is 100 oriented and $3\times3\times1.5$ mm$^3$ in size. The sample was implanted with 500 keV He ions at a fluence of $1\times10^{17}$ cm$^{-2}$ with a scanning ion microbeam of the 5U NEC Pelletron Accelerator of the University of Melbourne.

A variable thickness mask was deposited on the sample surface to control the ion penetration depth. The mask was obtained by means of Ag thermal deposition on the diamond surface through a hole. As shown in Fig. 7, the sample surface was tilted (>45°) with respect to the metal source so that the deposited mask was at least 2 µm thick in correspondence with its thickest part, while gradually decreasing in thickness down to a few nanometers. As reported in [19], implanting the specimen by scanning the ion beam across the mask allows the creation of an induced damaged region that emerges from the bulk (at a depth of about 0.4 µm) to the surface.

Transmission Electron Microscopy (TEM) cross-sections of the sample were realized to measure the thickness and depth of the amorphous carbon. The same above-mentioned FIB milling and lift-out procedure was used to prepare the sample lamellae. This process involved the deposition of a few micrometers of platinum in the area that was to be milled with the purpose of protecting the diamond top surface from unwanted FIB damage. Bright field TEM images were taken with the sample [110] zone axis parallel to the electron beam. The amorphous carbon layer has much lighter contrast in Fig. 8 due to its lower density and the elimination of diffracted beams from the crystalline diamond in the image formation by the objective aperture. Also, this layer appears uniform in intensity due to the absence of any diffraction contrast (thickness or bent contours) in amorphous materials. The significantly different intensity between amorphized regions and pristine diamond allows the detection of the rising amorphous carbon layer, as seen in Fig . 8.

The TEM image was processed and analyzed using Matlab to identify the positions of the edges of the a-C channel, as well as the interfaces between silver mask and diamond layer extending between the sample surface and the buried amorphous layer ("cap layer"). The image analysis allows the precise determination of the thickness and depth of the buried channel as a function of the horizontal spatial coordinate, as well as the corresponding thickness of the Ag mask, which varies from 0.52 µm to 0.90 µm. These thickness and depth values can be compared to the numerical values obtained by applying the model outlined in Section 2 to the output of TRIM simulations. The procedure was to first perform the TRIM simulation for 500 keV He ions implanted in the Ag mask (for the given thickness) and diamond substrate, then to apply model corrections to account for saturation and swelling, using the parameter values given above ($\alpha = 7 \times 10^{22}$ cm$^{-3}$ and $\rho_{aC} = 2.14$ g cm$^{-3}$), and finally to estimate the channel thickness by determining above-$\rho_{th}$ values in the resulting curve ($\rho_{th} = 4 \times 10^{22}$ cm$^{-3}$).

The experimental TEM values are compared to simulated values in Fig. 9, giving a very good agreement. Both datasets show a decrease in channel thickness for an increasing depth, with a 13% discrepancy between the slopes of the two linear fits on the data (not shown). This small discrepancy between experimental and simulated values can possibly be attributed to a small variation of $\rho_{th}$ , as in the case of $D_C$, with depth (at least when it is evaluated as a vacancy density as herein, and not as a mass density or strain), highlighted in previous experimental observations[15, 16, 25-33]. In particular, since in our model we are assuming a constant $\rho_{th}$ value as a function of depth, a possible systematic over-estimation of the layer thickness at greater depths (i.e. > 250 nm) would indeed imply that there is a slight increase of $\rho_{th}$ as a function of depth. In Fig. 9 there is no fully unequivocal evidence of such a trend in the mismatch between numerical and experimental data, therefore in the present work we can only formulate this observation as a hypothesis, which will need to be tested more strictly with data extending over broader depth ranges. Nonetheless, it should be stressed that the above procedure with a variable thickness mask is equivalent to verifying the predictive capabilities of the proposed model for implantation energies varying between 0 and 500 keV, thus confirming the robustness of the method.

**Conclusions**

We have introduced a modified and improved phenomenological model to account for damage accumulation and saturation in ion-implanted diamond, and combined it with FEM simulations to determine mechanical strains and surface deformations occurring in irradiated diamond for different ions and ion energies, both below and above its amorphization threshold. We have used various experimental techniques (HR-XRD, EELS, surface profilometry and TEM) to test the consistency of predicted numerical values. The agreement is found to be very good in most cases, and the presented analytical/numerical procedure proves to be a valuable predictive tool to determine internal strains and stresses in ion-implanted diamond. Since diamond amorphization has been found to be essentially strain-driven, this tool becomes essential when designing and performing high-accuracy microfabrication processes in this material with ion beam lithographic techniques.

**Acknowledgments**

The authors wish to thank Dr. Paolo Schina for the 180 keV B ion implantation at the Olivetti I-JET facilities of Arnad (Aosta, Italy). This work is supported by the FIRB "Futuro in Ricerca 2010" project "RBFR10UAUV", which is gratefully acknowledged. A.D.G. acknowledges the Australian Research Council for financial support (Project No DP0880466).

**List of Figure captions**

Fig. 1: Reciprocal space map measured with HR-XRD on the (0 0 4) reciprocal lattice point for the sample implanted with 180 keV B ions at a fluence of $5\times 10^{14}$ cm$^{-2}$. The colour scale indicates the diffracted intensity. The fringes along the $Q_z$ axis are an indication of the deformation gradient in the sample.

Fig. 2: Experimentally measured and simulated rocking curves from the sample implanted with 180 keV B ions at a fluence of $5\times 10^{14}$ cm$^{-2}$. The fringes appearing on the left side of the diffraction peak are fitted with a $10^{th}$ order spline, allowing the determination of the depth-dependent deformation profile that generates them.

Fig. 3: a) Experimentally derived relative lattice mismatch $\Delta d/d$ curves for the three samples implanted with 180 keV B ions (fluence values in the legend); b) $\Delta d/d$ curves, normalized with respect to an arbitrary factor proportional to their implantation fluence, for comparison purposes. The curves are superimposed on the linear vacancy depth profile $\lambda(z)$ calculated using the C-TRIM code, to highlight the correlation between deformation and vacancy density.

Fig. 4: Experimental (HR-XRD) and numerical (FEM) tensile strains as a function of depth for the sample implanted with 180 keV B at a fluence of $5\times10^{14}$cm$^{-2}$. The two datasets are in good agreement both in amplitude and depth distribution.

Fig. 5: a) Predicted vacancy density vs. depth profiles for 500 keV He implantations: TRIM simulation (red continuous line) and model predictions with only damage saturation (dashed green line) or both saturation and swelling (dotted blue line). b) Experimental and numerically predicted mass density vs. depth profiles for the same implantation. Additionally, a numerical curve (dashed orange line) accounting for "threshold" effects due to self-annealing is also reported. The figure graphically depicts how the model allows to determine an effective mass density variation in the material from the TRIM output.

Fig. 6: a) FEM - simulated $\varepsilon_z(x,z)$ strain field in the area implanted with 500 keV He at a fluence of $5\times10^{16}$ cm$^{-2}$, superimposed on the deformed profile (note: the vertical axis is magnified by a factor 50 with respect to the horizontal axis, to highlight the swelling effect); b) corresponding experimental and numerical surface swelling profiles.

Fig. 7: 3-D and 2-D views of the setup for the deposition of a variable-thickness mask through metal evaporation on a tilted (>45 deg.) diamond sample. The variable-thickness zone is highlighted. The mask is used for the creation of an emerging damage-layer after ion implantation.

Fig. 8: Bright field TEM image of an amorphous carbon layer rising to the surface in a diamond sample implanted with 500 keV He beam at fluence $1\times10^{17}$ cm$^{-2}$ through a variable-thickness Ag mask, which varies from 0.52 µm to 0.90 µm. The edges of the a-C region run parallel to the top surface of the mask, with only a slight thickness variation. The region analyzed in Fig. 9 ("processed region" is shown in the box.

Fig. 9: Experimental (continuous red line) and numerical (dashed blue line) thickness of the emerging a-C layer formed in diamond by 500 keV He ion implantation through a variable-thickness mask.

Figure 1

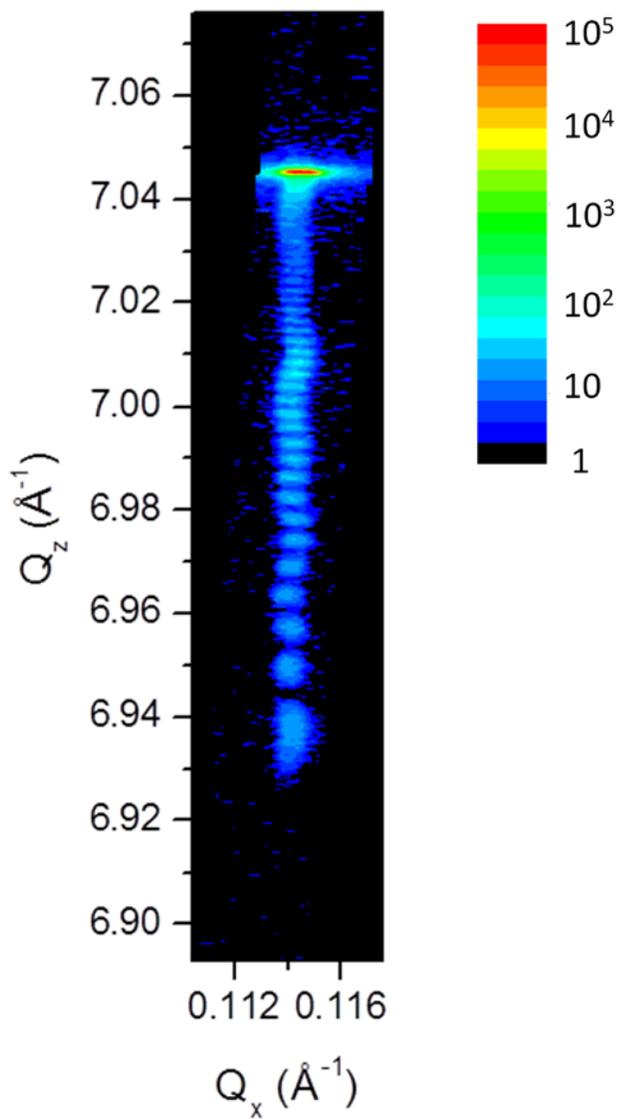

Figure 2

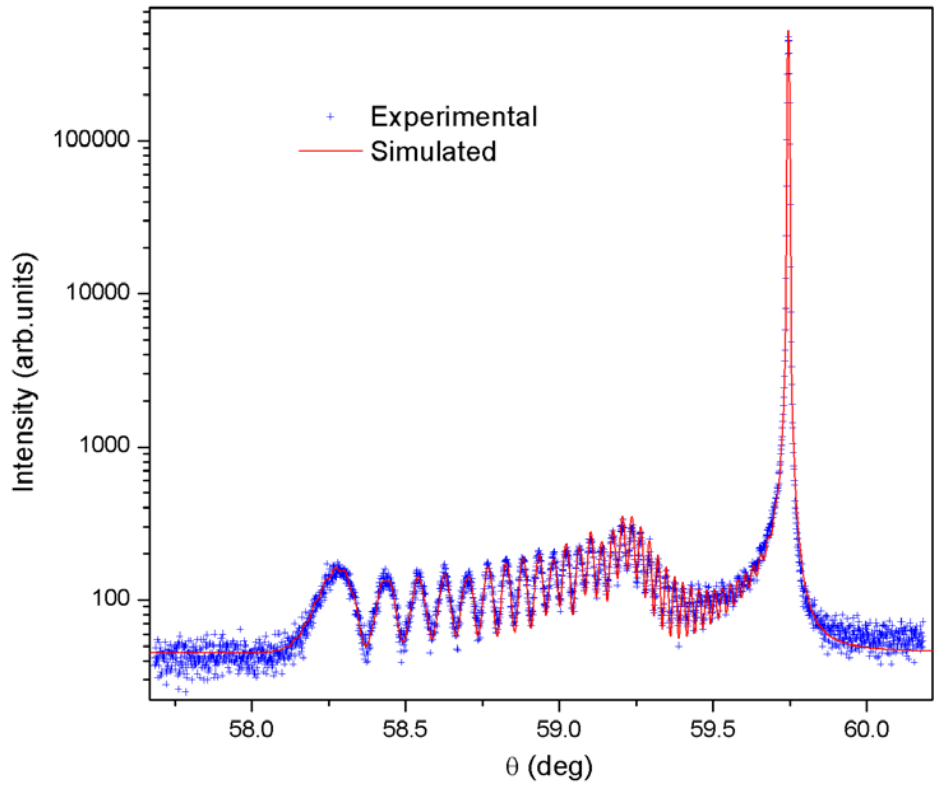

Figure 3

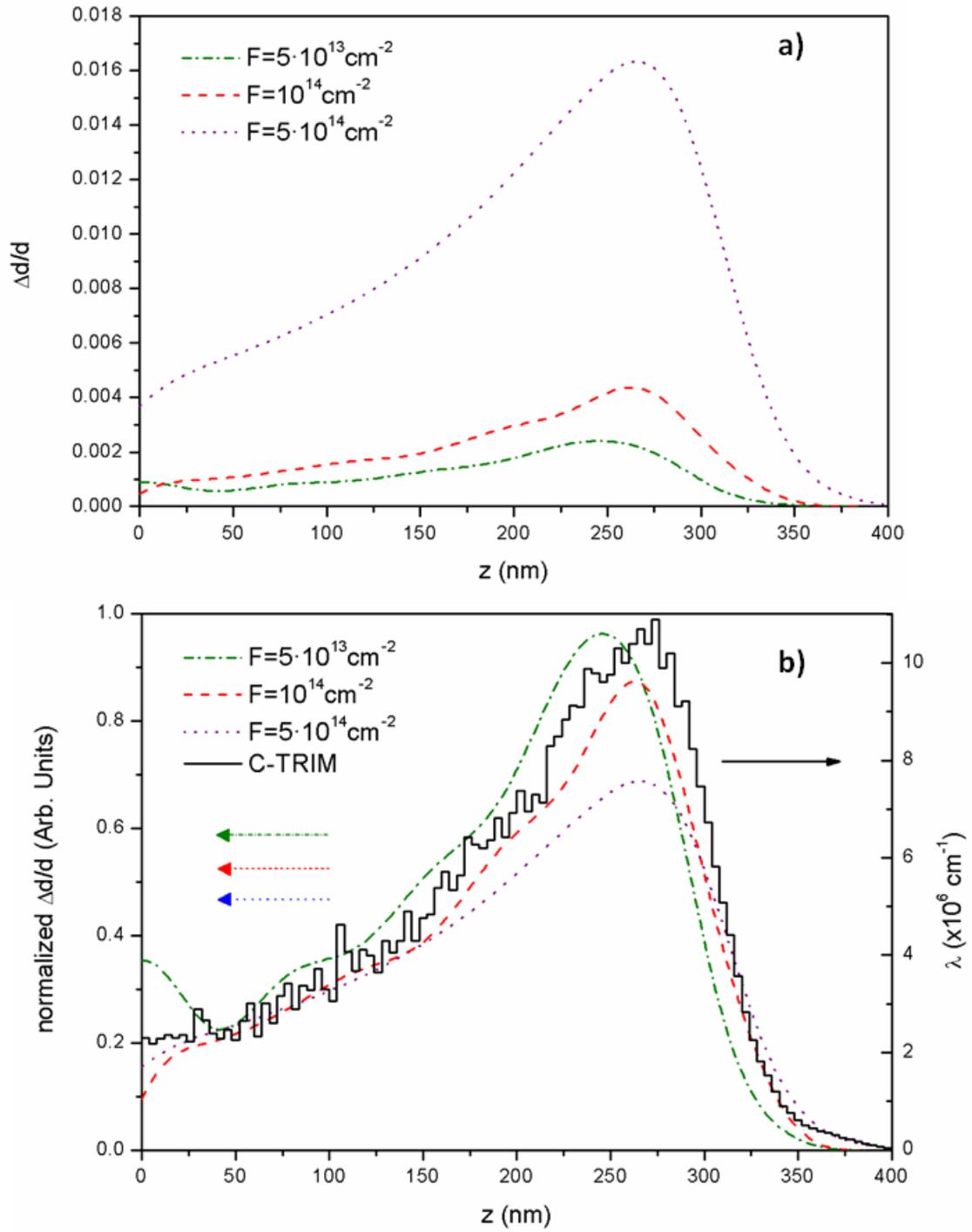

Figure 4

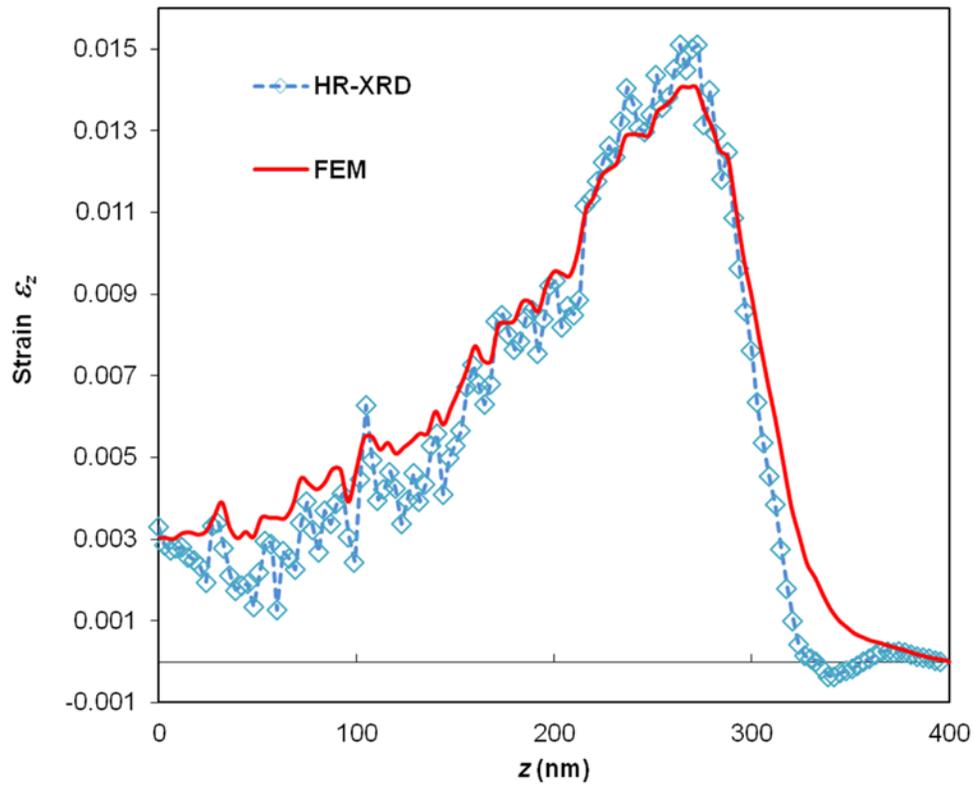

Figure 5

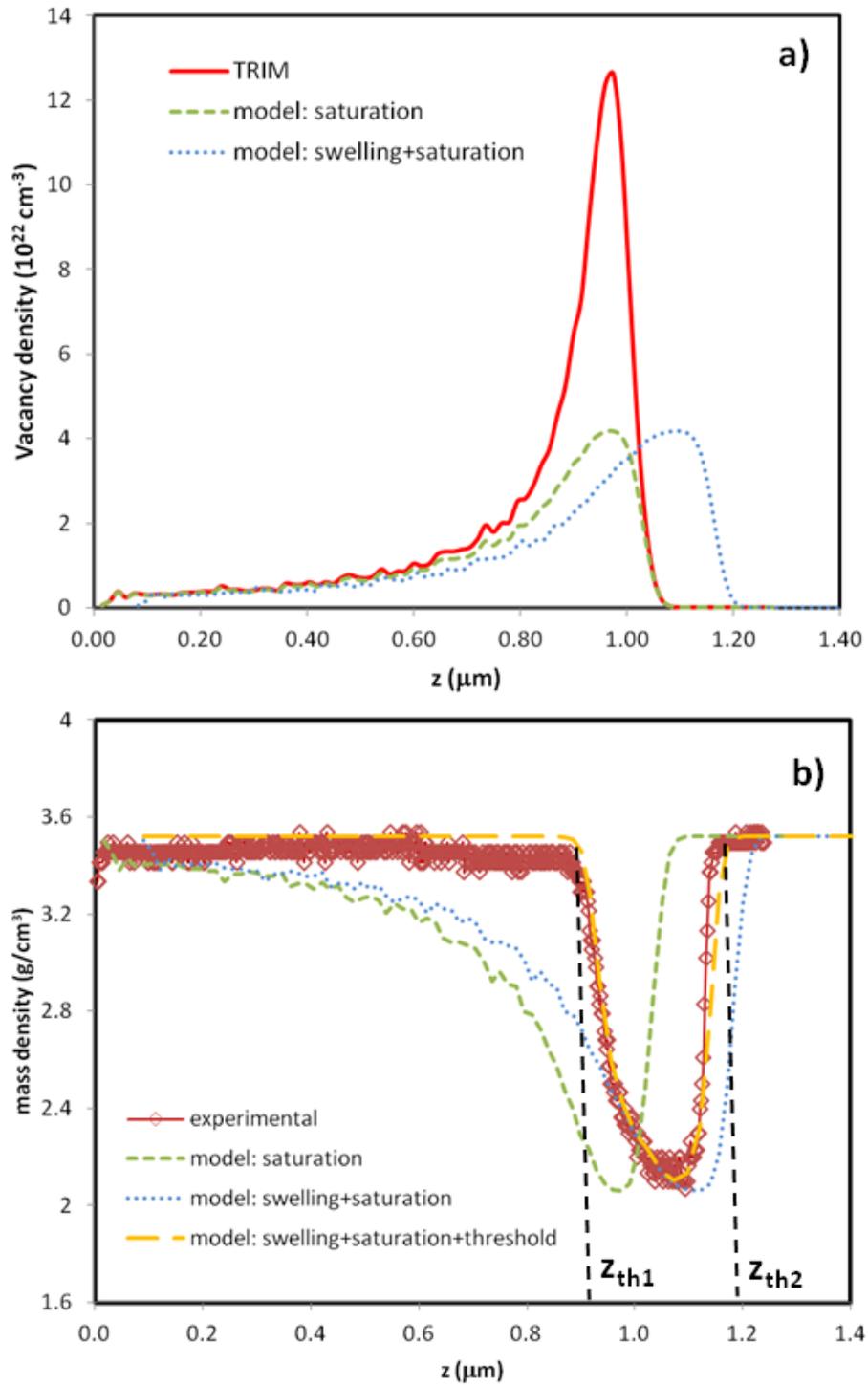

Figure 6

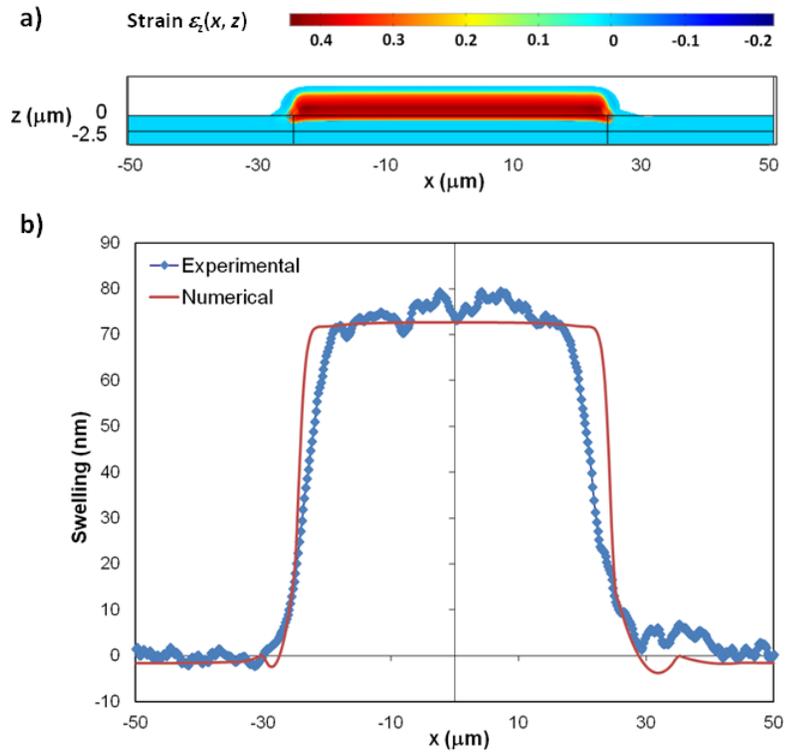

Figure 7

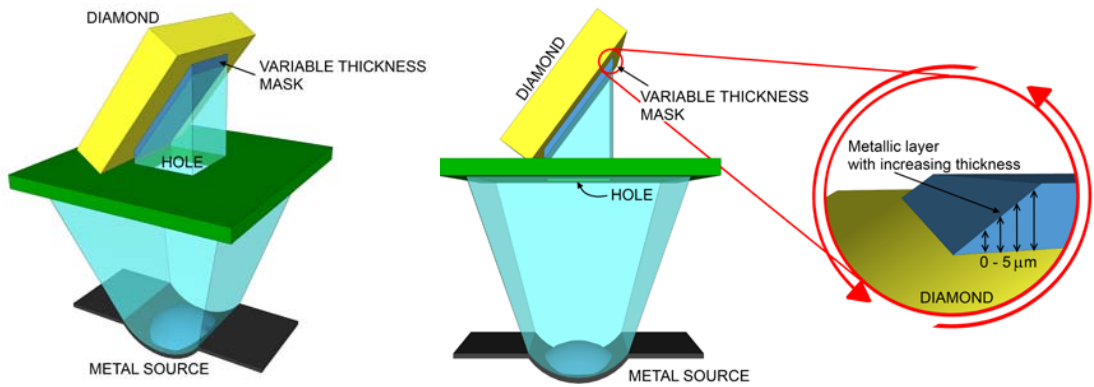

Figure 8

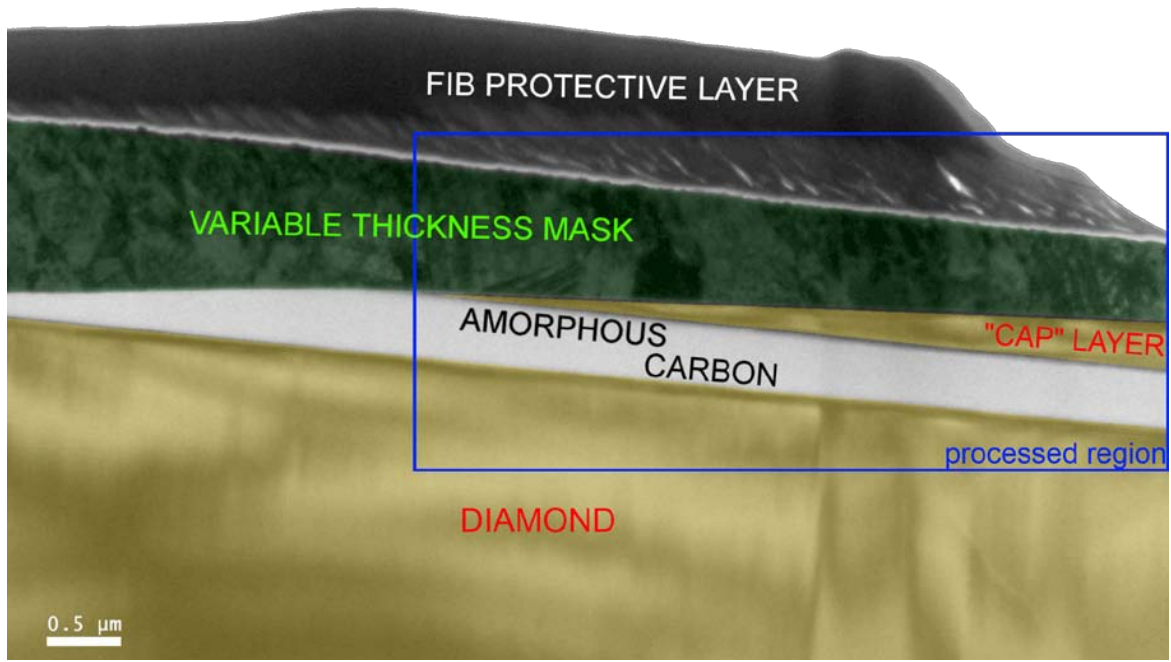

Figure 9

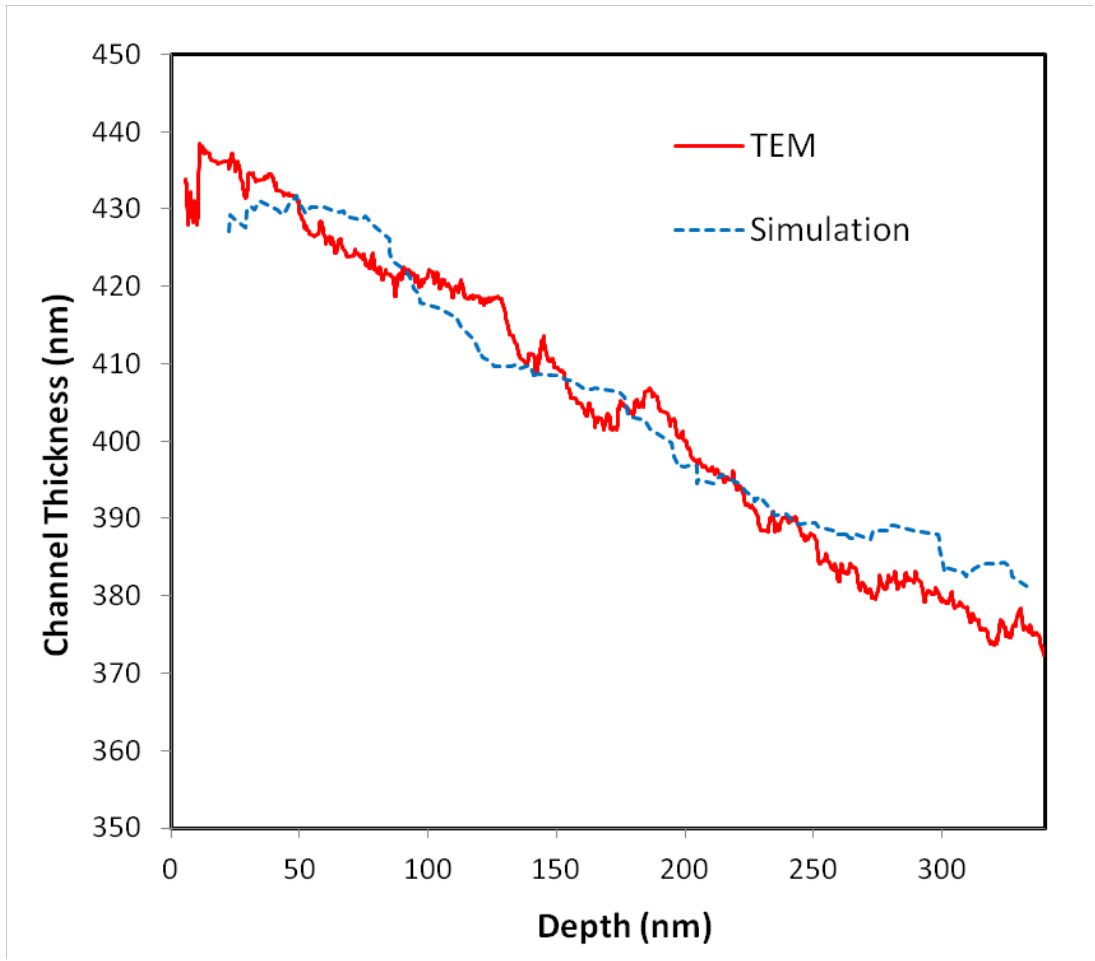